\documentclass[aps, prl, reprint,graphicx,groupedaddress,amsmath,amssymb]{revtex4-1} 
\newcommand{\JournalTitle}[1]{#1}
\usepackage{dcolumn}
\usepackage{bm}
\usepackage{natbib}
\usepackage{bibentry}
\usepackage[usenames,dvipsnames] {color}
\usepackage[table]{xcolor}
\usepackage{graphicx}
\usepackage{epstopdf}
\usepackage{amssymb}
\usepackage{amsmath}
\usepackage{dcolumn}
\usepackage{subfigure}
\usepackage{titlesec}
\usepackage{caption}

\DeclareGraphicsExtensions{.eps,.png,.pdf}
\DeclareMathOperator{\Tr}{Tr}
\usepackage{xr-hyper}
\usepackage{hyperref}
\hypersetup{
colorlinks   = true, 
urlcolor     = blue, 
linkcolor    = blue, 
citecolor   = blue 
}
\usepackage{setspace}
\usepackage[font=singlespacing,font=footnotesize, justification=RaggedRight]{caption}
\usepackage{enumitem}
\usepackage{fancybox}
\usepackage{comment}
\usepackage{url}
\usepackage{here}
\setlist{nolistsep}
\usepackage{multimedia}
\usepackage{soul}
\setstcolor{red}
\DeclareMathAlphabet      {\mathbf}{OT1}{cmr}{bx}{n}

\newcommand{\figref}[1]{Fig.~\ref{#1}}

\newcommand{\printfnsymbol}[1]{%
\textsuperscript{\@fnsymbol{#1}}%
}
\makeatother

\definecolor{gold_metallic}{rgb}{0.83, 0.69, 0.22}
\definecolor{frenchlilac}{rgb}{0.53, 0.38, 0.56}
\definecolor{indigo}{rgb}{0.0, 0.25, 0.42}
\definecolor{wildwatermelon}{rgb}{0.99, 0.42, 0.52}
\definecolor{wildstrawberry}{rgb}{1.0, 0.26, 0.64}
\definecolor{cadetblue}{rgb}{0.37, 0.62, 0.63}
\definecolor{caribbeangreen}{rgb}{0.0, 0.8, 0.6}

\graphicspath{{./Figures/}{./Movies/}}
\usepackage{siunitx}
\usepackage{gensymb}

\begin{document}

\title{Dynamics of active defects on the anisotropic surface of an ellipsoidal droplet}

\author{Martina Clairand$^{1}$}
\thanks{equal contribution}
\author{Ali Mozaffari$^{2,3}$}
\thanks{equal contribution}
\author{Jer\^ome Hardo\"uin$^{4,5}$}
\author{Rui Zhang$^{2,6}$} 
\author{Claire Dor\'e$^{1}$}
\author{Jordi Ign\'es-Mullol$^{4,5}$} 
\author{Francesc Sagu\'es$^{4,5}$} 
\author{Juan J. de Pablo$^{2,7}$} 
\email{depablo@uchicago.edu}
\author{Teresa Lopez-Leon$^{1}$} 
\email{teresa.lopez-leon@espci.fr}

\affiliation{$^1$ Laboratoire Gulliver, UMR CNRS 7083, ESPCI Paris, PSL Research University, 75005 Paris, France}
\affiliation{$^2$ Pritzker School of Molecular Engineering, The University of Chicago, Chicago, Illinois
 60637, USA}
 \affiliation{$^3$ OpenEye Scientific, Cadence Molecular Sciences, Boston, Massachusetts 02114, USA}
\affiliation{$^4$ Departament de Qu\'{\i}mica F\'{\i}sica, Universitat de Barcelona, 08028 Barcelona, Spain}
\affiliation{$^5$ Institute of Nanoscience and Nanotechnology, Universitat de Barcelona, 08028 Barcelona, Spain}
\affiliation{$^6$Department of Physics, Hong Kong University of Science and
  Technology, Clear Water Bay, Kowloon, Hong Kong}
\affiliation{$^7$Center for Molecular Engineering, Argonne National Laboratory, Lemont, Illinois
 60439, USA}
 
\date{\today}

\begin{abstract}
   Cells are fundamental building blocks of living organisms displaying an array of shapes, morphologies, and textures that encode specific functions and physical behaviors. Elucidating the rules of this code remains a challenge. In this work, we create biomimetic structural building blocks by coating ellipsoidal droplets of a smectic liquid crystal with a protein-based active cytoskeletal gel, thus obtaining core-shell structures. By exploiting the patterned texture and anisotropic shape of the smectic core, we were able to mold the complex nematodynamics of the interfacial active material and identify new time-dependent states where topological defects periodically oscillate between rotational and translational regimes.  Our nemato-hydrodynamic simulations of active nematics demonstrate that, beyond topology and activity, the dynamics of the active material are profoundly influenced by the local curvature and smectic texture of the droplet, as well as by external hydrodynamic forces.   
  These results illustrate how the incorporation of these constraints into active nematic shells orchestrates remarkable spatio-temporal motifs, offering critical new insights into biological processes and providing compelling prospects for designing bio-inspired micro-machines.
\end{abstract}

\maketitle

Active nematics (ANs), composed of elongated self-propelled units, have emerged as a paradigm to understand a range of biological processes, from the reorganization of active fibers during morphogenesis to the collective dynamics of bacterial populations and tissue growth \cite{marchetti2013, needleman2017, doostmohammadi2018, zhang2021auto}. By studying these systems under the umbrella of ANs, it becomes possible to exploit the theoretical framework that has been developed for equilibrium liquid crystals over the course of several decades to the profit of biology.\par 

In particular, nematic defects, singular points where the orientational order is disrupted, have been found to play a key role in the regulation of cellular function at both single-cell and multicellular levels \cite{maroudas2021, meacock2021, copenhagen2021, fardin2021}. However, the mechanism by which topological defects interact with their environment to produce functional outcomes remains an open question. \textit{In vitro} ANs provide a model platform to investigate this. Frictional forces \cite{thampi2014, doostmohammadi2016, thijssen2021, mozaffari2021, martinez-pratScalingRegimesActive2021a}, surface viscous anisotropy \cite{guillamat2016, guillamat2017, thijssenActiveNematicsAnisotropic2020}, spatial patterning of active stresses \cite{zhang2021spatio, tang2021, mozaffari2021, zhang2022logic}, or confinement \cite{wioland2013, ravnik2013, keber2014, zhang2016, khoromskaia2017, wu2017, ellis2018, guillamat2018, sokolov2019, pearce2019, opathalage2019, hardouin2020, rajabi2021} have recently been used to program defect motion and induce organized flows in the material, offering new fundamental insights and interesting perspectives for the design of bio-inspired micro-machines. However, the effect of substrate curvature, crucial in cellular processes, has been little explored experimentally, due to the challenges in producing controlled curved environments to probe ANs. \par 

In a pioneering study, Keber \emph{et~al.} \cite{keber2014} studied for the first time the effect of imposing a constant Gaussian curvature to a microtubule-kinesin AN, in which bundles of motor-propelled cytoskeletal filaments slide relative to each other, thereby producing a two-dimensional nematic field. In this material, a large number of topological defects randomly nucleate and annihilate, typically inducing a chaotic dynamics. By confining the AN to the surface of a spherical vesicle, they were able to stabilize a regular dynamic state characterized by the motion of four topologically required defects \cite{lubensky1992, lopez2011}. These experimental results have been replicated using various theoretical and numerical approaches \cite{zhang2020dynamics, brownTheoreticalPhaseDiagram2020}. Moreover, high activity conditions are predicted to lead to the emergence of novel structures, such as vortices \cite{sknepnek2015, khoromskaia2017, janssen2017, zhang2016} and stable rotating-bands \cite{sknepnek2015, henkes2018, castro2018}.




Imposing a gradient of Gaussian curvature is expected to induce spatio-temporal patterns of greater complexity\cite{alaimo2017, ehrig2017, pearce2019}. In curved geometries with non-uniform Gaussian curvature, such as an ellipsoidal or toroidal surface, curvature gradients act as external fields inducing defect attraction towards regions of like-sign Gaussian curvature or even defect unbinding under certain conditions \cite{bowick2009, bates2010, kralj2011, de2016, jesenek2015}. Such phenomena can dramatically influence the dynamics of the active system. For instance, when confining the AN to the surface of a prolate ellipsoid, curvature gradients are expected to trigger the segregation of defects to the poles \cite{bates2010, kralj2011} and the formation of tunable rotating states \cite{alaimo2017}. Although the interplay between non-uniform Gaussian curvature and defect dynamics has been experimentally explored for the first time using toroidal droplets, the sizes of these droplets were significantly larger than the active length scale that is proportional to the average distance between defects \cite{ellis2018}. As a result, the large number of defects that populated the droplet surface gave rise to chaotic regimes, and there is no experimental evidence of the regular dynamics predicted, particularly for the case of the ellipsoidal geometry. \par

In this work, we take advantage of the physical properties of smectic liquid crystals to fabricate ellipsoidal droplets with dimensions comparable to the typical length scale of the AN. By introducing these elongated droplets in a microtubule-kinesin active bath, we are able to stabilise an ellipsoidal AN layer at the aqueous/smectic interface of the droplet. In this configuration, we witness the emergence of two new time-dependent dynamic states, with dipolar and quadrupolar symmetry respectively, whose dynamics is controlled by the periodic motion of two pairs of topologically required $+1/2$ defects. In both, the quadrupolar and the dipolar state, the system oscillates between two different regimes: i) a rotational regime, where defects rotate around the major axis of the ellipsoid, forming swirls at the poles and, under certain conditions, an equatorial rotating band, and ii) a translational regime, triggered by bending instabilities, where two pairs of $+1/2$ defects commute from one pole to the other one, aligning the active flows longitudinally. The transition from chaotic to well-ordered modes is discussed in terms of a characteristic length scale defined by the balance between active and elastic stresses in constrained systems \cite{hemingway2016}. The prevalence of a final dipolar state is supported by nemato-hydrodynamic simulations that reconstitute the dynamics observed experimentally. While the main features of the defect dynamics result from the imposed ellipsoidal geometry, the preservation of chirality over time stems from the interplay between viscous anisotropy, enforced by the smectic substrate, and hydrodynamics. These findings provide a new light for understanding some biological processes and for the design of bio-inspired micro-machines.

\section* {Building an ellipsoidal active nematic}
\subsubsection*{Smectic ellipsoidal droplets}

Our ellipsoidal ANs consist of elongated droplets of octyl-cyanobiphenyl (8CB), a passive 
liquid crystal, coated by a layer of aligned microtubule bundles sheared by kinesin motors. 

The fabrication of the smectic droplets relies on the temperature-induced bursting of 
water/8CB/water double emulsions. The latter are produced in a glass capillary microfluidic device \cite{utadaMonodisperseDoubleEmulsions2005a}, using an aqueous solution of Pluronic F-127 (P2443, Sigma) ($2\;\rm{wt.}\%$) as inner 
phase, 8CB as middle phase, and an aqueous mixture of Pluronic F-127 ($2\;\rm{wt.}\%$) 
and glycerol ($60\;\rm{wt.}\%$) as outer phase. More details on the protocol to produce 
the double emulsions are available in the \hyperref[sec:material]{Materials and Methods} section. 

To obtain the ellipsoidal shape, the double emulsions must be produced and collected at 
a temperature ranging between \SI{33.5}{\celsius} and \SI{40.5}{\celsius}, so that 8CB is 
in the nematic phase. Upon quenching the system, a nematic-smectic phase transition is triggered, causing the destabilization and collapse of the double emulsions into ellipsoidal droplets of 8CB. This finding is noteworthy, because 8CB droplets produced by 
microfluidics or vortexing typically adopt a spherical shape. The molecular organization of the system in the shell geometry, prior to the formation of the droplet, seems to be the reason behind the exotic shape of the droplets generated with this method. The details of this transformation are the subject of an independent study. The resulting smectic ellipsoids are metastable, and tend to become spherical after approximately 48 hours, yet their elongated shape remains stable enough to conduct experiments on the active system. For this study, we have considered ellipsoidal droplets with principal semi-axes of length $a=b=(25-45) \rm{\mu m}$ and $c=(50-90) \rm{\mu m}$, choosing those with an aspect ratio of $c/a \approx 2$.

\begin{figure*}[!htb]
 \centering
 \includegraphics[width=.9\linewidth]{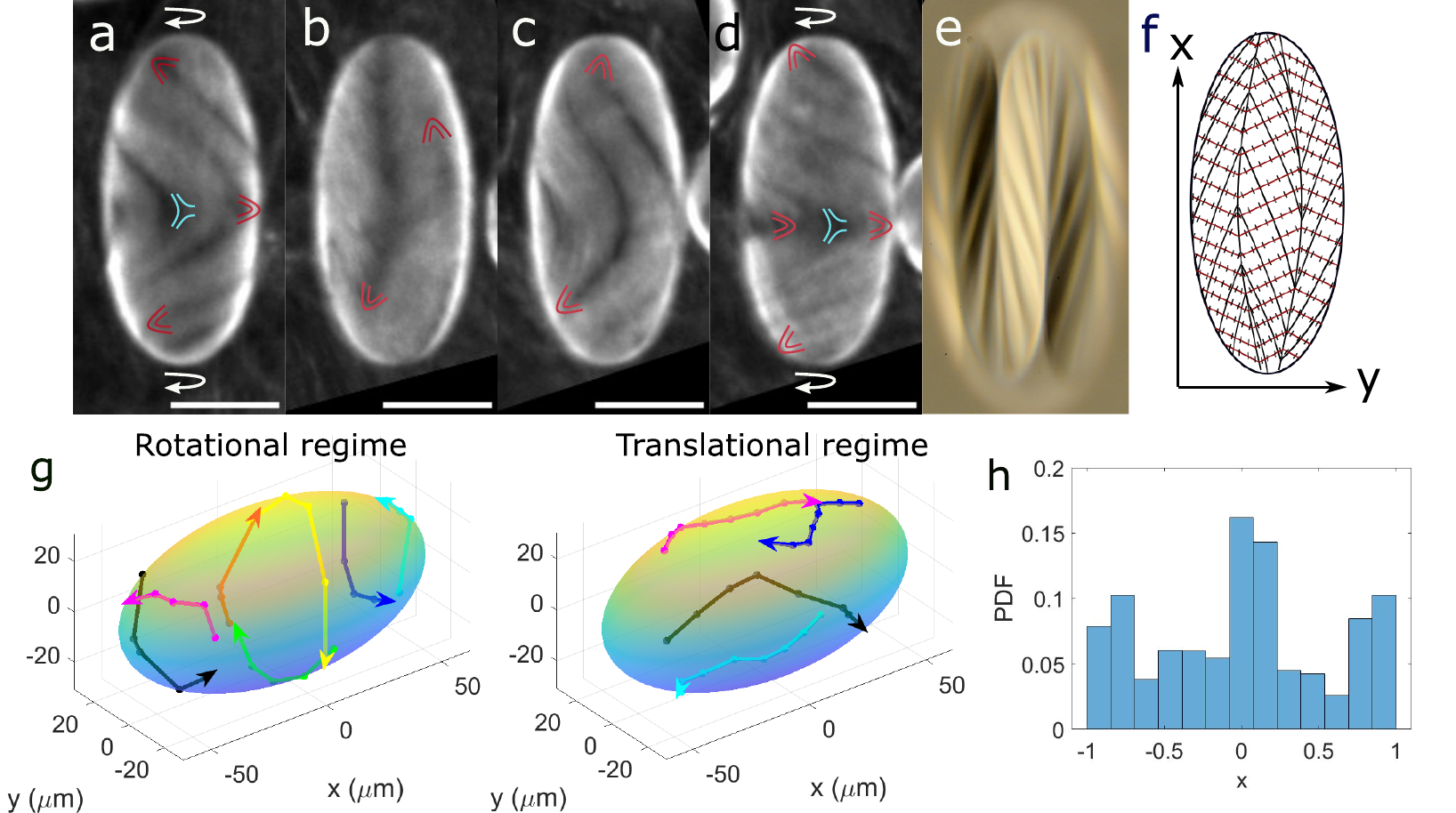}
 \caption{Defect dynamics in the transient quadrupolar dynamic state. (a)-(d) Fluorescence confocal micrographs showing half-period of the system dynamics, from $t=0$ (a) to $t=23\,\rm{s}$ (d).  Parabolic (red) and hyperbolic (cyan) lines indicate $+1/2$ and $-1/2$ defects, respectively. Scale bars are $50 \mu m$. (e) Crossed polarized micrograph of a smectic ellipsoidal droplet. (f) Schematic representation of the director field in (e). The smectic layers (red lines) display a chevron pattern and are perpendicular to the director field (black lines). (g) 3D reconstruction of the trajectory of $\pm 1/2$ defects in the rotational and translational regimes in (a)-(d). (h) Normalized probability distribution histogram for the position of $\pm 1/2$ defects along the major axis of the ellipsoid. The dynamics reported in (a)-(d) were observed for 50 droplets in different samples.  The trend in (h) was observed for all the analyzed droplets (typically 10).}
 \label{fig_1}
\end{figure*}

Under both bright field and polarized light, the smectic ellipsoids reveal a structure similar 
to that described for spherical smectic shells \cite{lopez2011nematic}. The most outstanding 
feature of their optical texture is the presence of a set of meridional lines, which connect 
the two poles of the ellipsoid, dividing it into crescent domains, as shown in \figref{fig_1}(e)-(f). 
These lines represent curvature walls where the director $\mathbf{n}$ rotates by a constant angle. Consequently, the smectic layers, which are indicated by red lines in \figref{fig_1}(f), are tilted in opposite directions in adjacent domains.

\subsubsection*{Adding an active nematic shell}

To produce the ellipsoidal AN, the smectic droplets are dispersed in an aqueous active gel, where energy consuming kinesin motors cross-link and set into motion adjacent fluorescent-labeled microtubule bundles. In contrast to previous studies, where the active gel was encapsulated inside a droplet or vesicle \cite{keber2014, ellis2018}, here the active gel constitutes the continuous phase in which the droplets are dispersed. Through the depleting action of polyethylene glycol (PEG), the bundles gradually condense at the aqueous/8CB interface to progressively form an ellipsoidal AN \cite{sanchez2012, decamp2015}. 

The active gel containing the passive droplets is introduced into a polyacrylamide-coated square glass capillary of $0.6$ mm inner width for observation. A constant supply of adenosine triphosphate (ATP) is ensured by the addition of phosphoenolpyruvate (PEP), which regenerates the activity of the nematic film. The chemical energy in the bulk can be renewed as often as needed through the dispersing solution, right before the activity is exhausted. An extended version of the protocol is provided in the \hyperref[sec:material]{Materials and Methods} section. \par

The dynamic self-assembly of microtubules and motor clusters can be observed a few minutes after filling the capillary with the active mixture. The dispersing phase is populated with fluorescent 
filaments, tens of microns long, that continually extend and buckle, generating large active 
chaotic flows around the passive droplets. Initially, only a small portion of the microtubule bundles that contact the ellipsoidal surface is adsorbed. Their progressive accumulation on the droplet surface with time makes the active network stiffer, driving the active length scale to higher values \cite{martinez2019, kampmannControllingAdsorptionSemiflexible2013, stepanowAdsorptionSemiflexiblePolymer2001}. Our observations of approximately 50 droplets reveal two distinct states, the early quadrupolar state and the final dipolar state, both characterized by well-defined nematic order, large-scale flows, and remarkable periodic dynamics.
The two states are described in the following section. \par


\section{Dynamic states of an active nematic ellipsoidal shell}\label{exp}

\subsubsection*{Quadrupolar state: The transient dynamic state}
After approximately two hours of the experiment, under the given experimental conditions, the droplet surface has accumulated enough active material to sustain an organized state characterized by an ordered pattern of textures and flows. The dynamics of the system exhibits clear switching behavior, as indicated by the motion of defects.  Specifically, the system commutes regularly between a long-lived regime of rotationally-moving defects that is interspersed with episodes where defects feature translational motion, as shown in \href{run:../SI/Movie_1.avi}{Movie 1} in SI. A half-period of this dynamics is reproduced in \figref{fig_1} (a)-(d). \figref{fig_1}(g) shows the trajectories described by the defects during a certain representative time for both the rotational and the translational regimes. 

\figref{fig_1}(a) and (d) show two snapshots of the droplet in the \textit{rotational regime}. In this regime, each pole of the ellipsoid hosts a pair of $+1/2$ defects, while the ellipsoid equator is decorated by a belt of $+1/2$ and $-1/2$ defects. Positive (self-propelled) and negative (dragged) defects are highlighted in red and cyan, respectively, in \figref{fig_1}(a) and (d). It should be noted that only the defects on the visible side of the ellipsoid appear in these images, although full trajectories for all defects are represented in \figref{fig_1}(g). 

At the poles, the defects follow quasi iso-latitude trajectories, see \figref{fig_1}(g) (left), while they maximize their separation distance due to elastic repulsion, as it will be discussed later. These defects turn about the ellipsoid major axis assembling $co-rotating$ swirls of the same handedness at the two poles of the ellipsoid, see \href{run:../SI/Movie_1.avi}{Movie 1} in SI. The chirality of this motion is determined by spontaneous symmetry breaking and is randomly established when the AN is formed, remaining conserved over time.

To conciliate the defect dynamics at the poles, a $counter-rotating$ defect belt appears simultaneously at the waist of the ellipsoid, setting a quadrupolar configuration of globally compensated torques. In the belt, oppositely charged defects form balanced pairs that do not contribute to the global topological charge of the system, which is $+2$, consistently with the global topological constraints for the ellipsoid \cite{hopfVektorfelderNdimensionalenMannigfaltigkeiten1927, poincareMemoireCourbesDefinies1881}. The optical texture obtained by fluorescence imaging shows the formation of transversal bands (nearly perpendicular to major ellipsoidal axis) displaying mirror symmetry with respect to  equatorial plane, see \figref{fig_1}(a) and (d). This pattern is compatible with the smectic structure of the substrate, shown in \figref{fig_1}(e) and (f), with the smectic layers setting the easy flow directions \cite{guillamat2017, thijssenActiveNematicsAnisotropic2020}, see section 2 in SI for more details.

Because extensile ANs are intrinsically unstable against bend deformations, such a transversally aligned state is susceptible to instabilities. This instability results in the appearance of four crimps perpendicular to the aligned field, with two visible in \figref{fig_1}(b) as dark longitudinal lines, while the other two are on the other side the droplet. This instability marks the onset of what we call the \textit{translational regime}, where the two pairs of $+1/2$ defects concurrently migrate along the crimps from pole to pole, locally reorienting the director field along the major axis of the ellipsoid, see \figref{fig_1}(b), \figref{fig_1}(g)(right) and \href{run:../SI/Movie_1.avi}{Movie 1} in SI. As a result, the transversal bands get reoriented into longitudinal bands. The longitudinally aligned configuration is also unstable against bending, and the straight bands rapidly adopt parabolic shapes displaying mirror symmetry with respect to the equatorial plane, see \figref{fig_1}(c). The equatorial belt with its accompanying population of positive and negative defects is progressively reconstituted, and the transversal bands reappear, see \figref{fig_1}(d). This configuration is the same one shown in \figref{fig_1}(b) with the pole defects having exchanged location. In the second half of the period, the whole process is repeated, and the two pairs of $+1/2$ defects regain their initial location. This incessant switching between rotationally and translationally moving phases persists for up to nearly two more hours, during which the active material continues to accumulate at the surface of the ellipsoid.

To investigate the effect of curvature on the defect dynamics, we analyzed the positioning of the defects in terms of their probability distribution along the major axis of the ellipsoid, see \figref{fig_1}(h). The histogram, which includes both positive and negative defects, aligns with two observations: firstly, positive defects tend to cluster around regions of maximal positive Gaussian curvature at the poles, and secondly, pairs of positive and negative defects with zero net topological charge persist in circulating along the equator. It is worth noting that this quadrupolar configuration is a transient state that eventually gives way to a more robust flow organization of a different symmetry, as described below.

 \par

\subsubsection*{Dipolar state: The final dynamic state}

\begin{figure*}[!htb]
 \centering
 \includegraphics[width=1.0\linewidth]{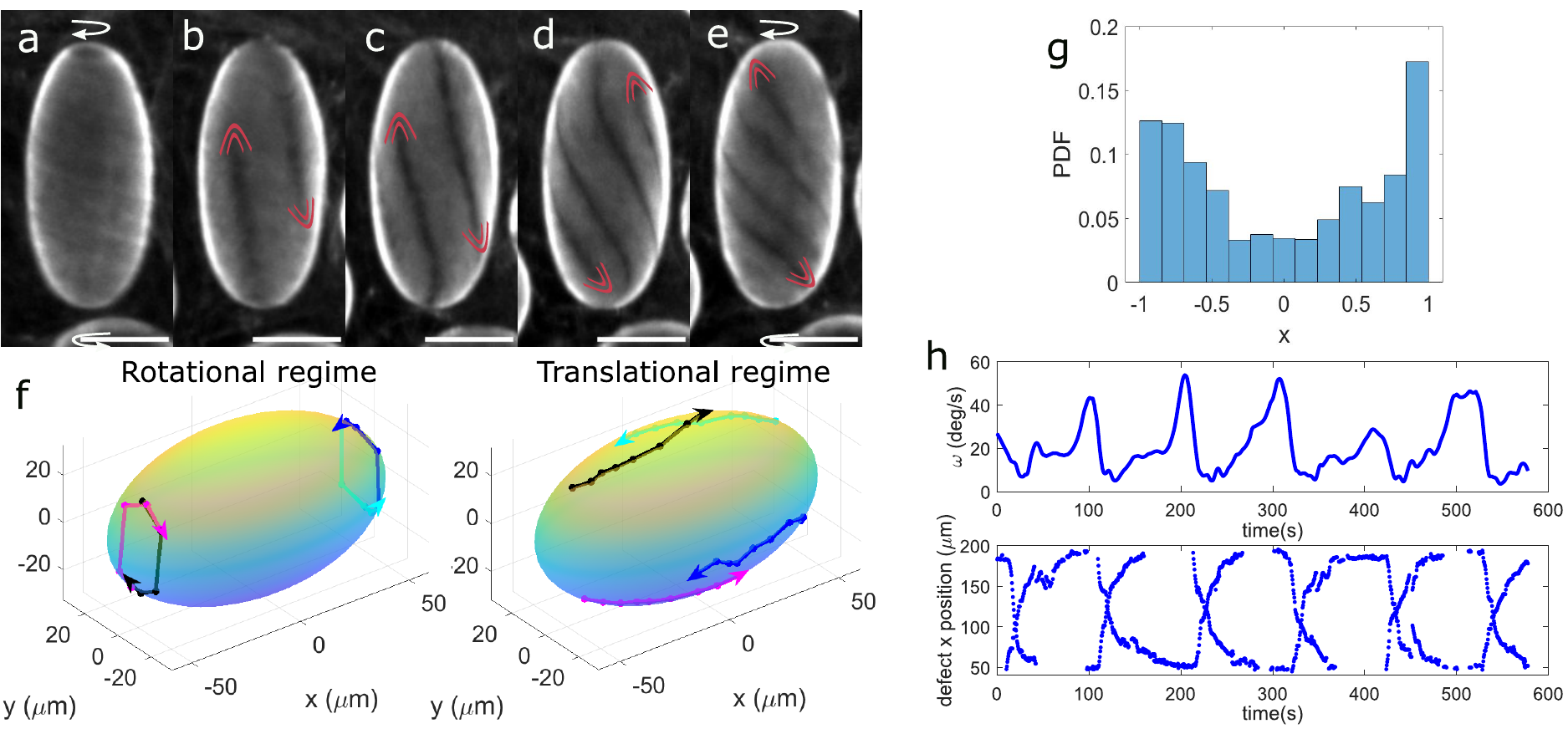}
 \caption{Defects dynamics in the final dipolar dynamic state. (a)-(e)Fluorescence confocal micrographs showing half-period of the system dynamics, from $t=0$ (a) to $t=15\,\rm{s}$ (e). Scale bars are $50 \mu m$. Parabolic red lines indicate the position of $+1/2$ defects. (f) 3D reconstruction of the $+1/2$ defect trajectories in the rotational and translational regimes. (g) Normalized probability distribution histogram for the position of $+1/2$ defects along the major axis of the ellipsoid. (h) Temporal evolution of the average angular speed and defect position along the major axis of the ellipsoid. The dynamics reported in (a)-(e) was observed for 50 droplets in different samples. The distribution trend in (g) and the periodic oscillations in (h) were observed for all the analyzed droplets (typically 10).}
 \label{fig_2}
\end{figure*}

As time progresses, the amount of active material adsorbed on the ellipsoid surface appears to reach saturation. This likely makes the active material stiffer, thereby stabilizing the textures and flows at larger length scales. This results in the suppression of all non topologically required defects, and thus, in an asymptotic dynamic state regulated by the motion of only two pairs of $+1/2$ defects. This final state displays some similarities and differences with the transient quadrupolar state described earlier. The most noticeable similarity is that the dynamics periodically switches between a rotational regime, where the two pairs of $+1/2$ defects form swirls at the ellipsoid poles, and a translational regime, where the defects commute from pole to pole. The most important difference is that, during the rotational regime, the system displays dipolar rather than quadrupolar symmetry. A half-period of this new oscillatory dynamics is shown in \figref{fig_2}(a)-(e), and representative defect trajectories for the rotational and translational regimes are plotted in \figref{fig_2}(f).

The \textit{rotational regime}, is characterized by the \textit{counter-rotation} of the two pairs of $+1/2$ defects around the poles, displaying opposite chirality, as depicted in \figref{fig_2}(a), \figref{fig_2}(f)(left) and \href{run:../SI/Movie_2.avi}{Movie 2} in SI. This is in contrast to the quadrupolar state, where the defects \textit{co-rotate} at the poles. In this \textit{counter-rotating} configuration, there is no need for a defect belt at the equator to compensate the flow dynamics at the poles. This arrangement conforms to a dipolar structure, with loss of equatorial mirror symmetry,  where the flows are well aligned along latitudinal lines, as shown in \figref{fig_2}(a). 

Bend instabilities act again to relax the accumulated stress on this aligned configuration, setting the onset of the \textit{translational regime}, where longitudinal flows trigger the migration of defects towards opposite poles, as shown in \figref{fig_2}(b)-(e) and \figref{fig_2}(f)(right). Unlike in the quadrupolar state, where the straight longitudinal bands bend into a parabolic shape, the longitudinal bands here spontaneously tilt and eventually re-form into two swirls at the poles, completing the first half of the dynamics, see \figref{fig_2}(b)-(e). In the second half, the entire process is repeated, and the defect pairs return to their initial positions. The chirality of the flows is preserved throughout the process. To visualize the defect trajectory over several dynamical cycles see \href{run:../SI/Movie_2.avi}{Movie 3} and the left panel of \href{run:../SI/Movie_2.avi}{Movie 4} in SI.  

The positioning of defects along the major axis of the ellipsoid is shown in the histogram of \figref{fig_2}(g), indicating that the defects preferentially accumulate at the poles where the Gaussian curvature is maximal, in agreement with theoretical predictions. \figref{fig_2}(g) shows the average angular speed of defects as a function of their position along the major axis of the ellipsoid. A pattern of in-phase oscillations is clearly evidenced, where the maximum values of the angular velocity correspond to the rotational regime,  and minimum values correspond to the translational regime, where the defects commute between poles. Additional details on the periodic oscillation of topological defects are presented in the SI, including the angular distances between defects as shown in Fig.S1 and Movie 4 in SI. \par


In summary, our findings demonstarte the existence of two oscillatory dynamic states in ANs confined to the ellipsoidal surface of a smectic droplet. Both gradients of Gaussian curvature and viscous anisotropy seem to control the active flows, leading to remarkably organized dynamic patterns. In the following section, we resort to numerical simulations to try to decouple the role of these two effects, otherwise impossible to disentangle in experiments.

\begin{figure*}[!htb]
 \centering
 \includegraphics[width=1\textwidth]{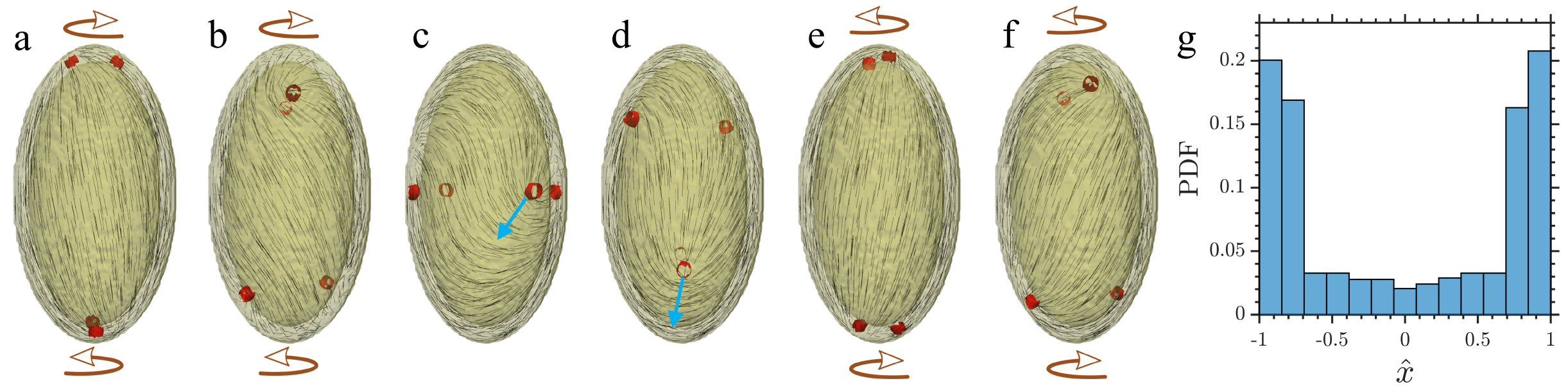}
 \caption{(a)-(f) Simulation results for the spatio-temporal evolution of the nematic texture of an AN shell
 confined between two concentric ellipsoids, shown in a series of snapshots. The
sense of rotation of defects at the poles is denoted by brown arrows. The red
 isosurfaces indicate the location of the four $+1/2$ defects and the black lines are the local
 director field. The blue arrow shows the direction of self-propulsion of the $+1/2$ defect. 
 (g) Probability distribution histogram for the location of the defects along the ellipsoid major axis. The defect position is normalized with the average of the inner and outer shell major axes.}
 \label{fig_3}
\end{figure*}


\section{Coupling between curvature gradients, friction anisotropy, and hydrodynamic flows}

\subsubsection*{Curvature effects}

The final dipolar state observed experimentally is simulated by a thin shell of AN in a tangent plane between two concentric ellipsoids. Initially, we neglect the effects of the underlying smectic layers and focus on the impact of curvature gradients on the trajectories and dynamics of defects. \par

In the absence of active stresses, each pole is occupied by a pair of $+1/2$ defects, with the director field tangentially aligned along the major axis of the ellipsoid to minimize elastic free energy. This configuration allows for a nearly uniform alignment of the director field in regions with low curvature. The active stresses bring the system out of this equilibrium configuration, and the defects start to rotate indefinitely on the antipodal points of a circumference centered at each pole of the ellipsoid, see \href{run:../SI/Movie_3_manu19.mp4}{Movie 5} in SI for a visual representation of this process. Away from the poles, the director field adopts a steady configuration with a slight tilt relative to the major axis. As the activity strength increases, the defects rotate faster, and the director field tilts further from the major axis. Similar to our experimental observations, when the activity surpasses a critical value, the defects display spiral trajectories and migrate toward opposite poles. The system develops well-organized spatio-temporal patterns, with defect dynamics that periodically switch from a circulation-dominant motion at the regions of maximal curvature, to linear translation near the equatorial plane, see \href{run:../SI/Movie_4_manu21.mp4}{Movie 6} in SI. Note that defects maintain their sense of rotation as they travel towards the opposite side of the ellipsoid, resulting in an inversion of the handedness at the poles every half-period, as shown in the sequence of images in \figref{fig_3}(a)-(f). We further explore the residence time of the defects through a probability  distribution function for finding defects along the major axis of the active system. \figref{fig_3}(g) reveals that defects are primarily found in the regions of maximum Gaussian curvature. At the higher values of activity, the residence time at the poles decreases, and the histogram shown in \figref{fig_3} becomes more uniform. \par

The dynamics described in these simulations, which we refer to as \textit{frictionless system simulations}, reproduce the periodic commutation between the rotational and translational regimes. However, in contrast with experiments, there is an inversion of the handedness at the poles every half period. Therefore, the experimental final state shown in \figref{fig_2} cannot be explained solely on the grounds of geometric considerations, and other parameters involved in the steering of the defects' pathways must be examined. Indeed, the smectic droplet under the AN shell displays a textured surface. As discussed in the Supporting Information, we believe that the complex dynamics of the active system and the trajectories of the self-propelled $+1/2$ defects are influenced by the anisotropic friction enforced by the smectic  substrate. To address this question, in the next section, we consider the effect of anisotropic friction in our simulations.

\begin{figure*}[!htb]
 \centering
 \includegraphics[width=1\textwidth]{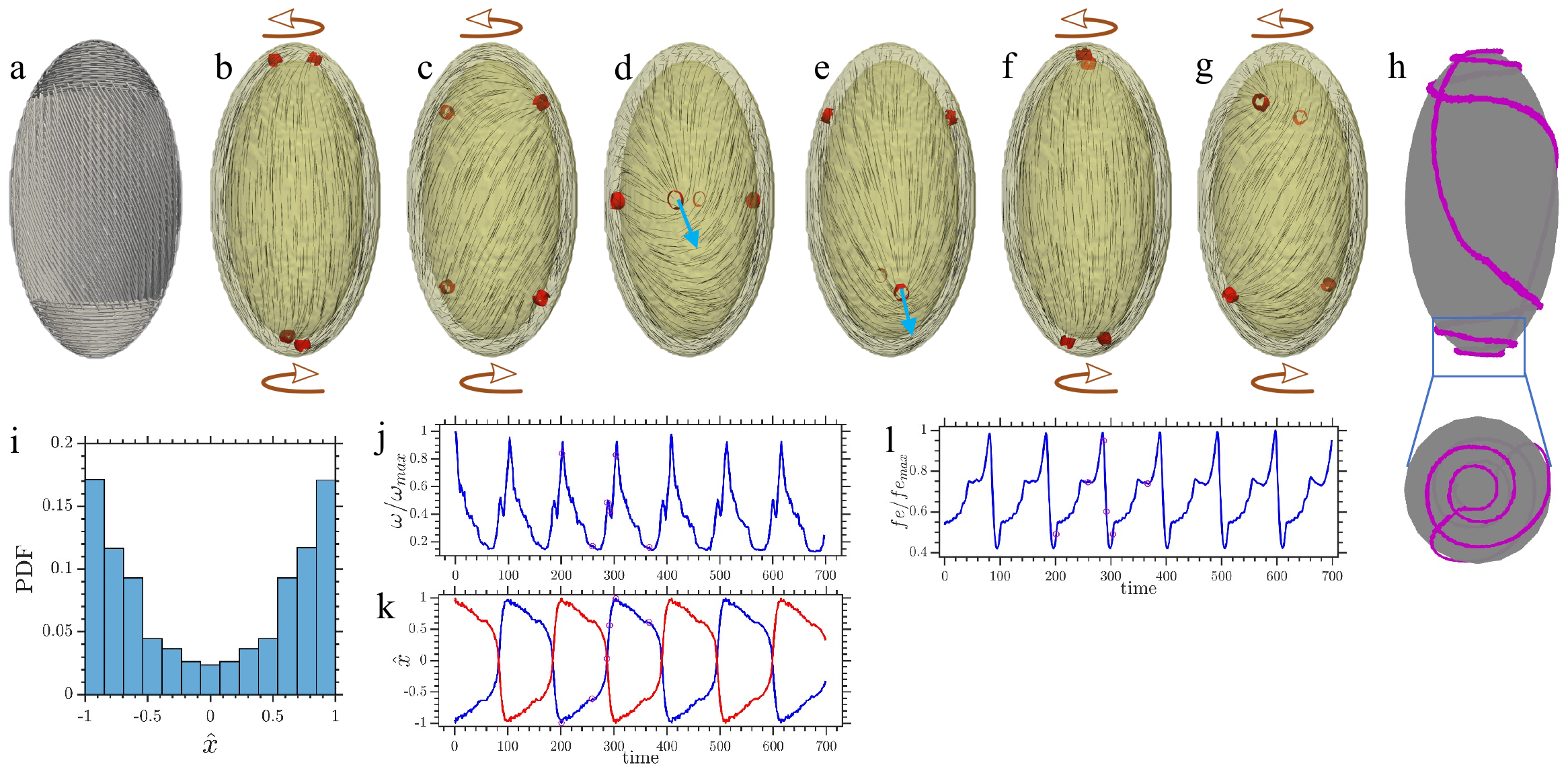}
 \caption{(a) Easy flow pattern resulting from the anisotropic frictional damping used in this simulation. (b)-(g) Evolution of the nematic director (black lines) and the position of defects (red isosurfaces), with arrows indicating the direction of circulation at the poles. The direction of motion of $+1/2$ defect is indicated by a blue arrow. (h) Side and top view of a single defect trajectory, which shows 
 the spiraling motion at the poles and a linear translation away from the poles. 
 (i) Probability distribution histogram for the position of defects along the long axis of the
 ellipsoid. (j) Time evolution of the magnitude of average angular velocities of defects.
 (k) Normalized position of two defects propelling from opposite poles of the ellipsoid as a function of
 simulation time. (l) Variation of elastic the free energy of the system.  The six purple open
 circles marked on the plots in \figref{fig_4}(j)-(l) correspond to the snapshots of
 \figref{fig_4}(b)-(g), respectively. Positions are normalized with the average of the major
 axes of the inner and outer shells.}
 \label{fig_4}
\end{figure*}

\subsubsection*{Role of friction anisotropy}

The effect of the smectic layers in contact with the AN is accounted for by adding a frictional damping force to the equations of motion that penalizes the fluid flow along certain directions, with a strength defined through a diagonal tensor $\bf{f}$. More details about this approach can be found in Section 2 of SI, where we discuss the role of frictional damping in the regulation of planar active flows. We first consider a uniform frictional pattern that penalizes flow along the long axis of the shell, aiming to mimic the structure of smectic layers, which provide an easy flow direction roughly parallel to the equatorial plane, see red lines in \figref{fig_1}(f). In agreement with experiments, we observe topological defects organized in pairs at the poles, $counter-rotating$ about the major axis of the ellipsoid. As defects spin around each other, the director field gets slightly tilted away from its original orientation. This simulation shows that the incorporation of anisotropic frictional damping forces improves agreement with the residence time distribution of the defects along the long axis observed in experiments, but fails to accurately predict the defects' trajectories and, even more importantly, the handedness of rotation of the defects at the poles, see \href{run:../SI/Movie_5_manu25.mp4}{Movie 7} in SI.\par

\begin{figure*}[!htb]
 \centering
 \includegraphics[width=0.9\textwidth]{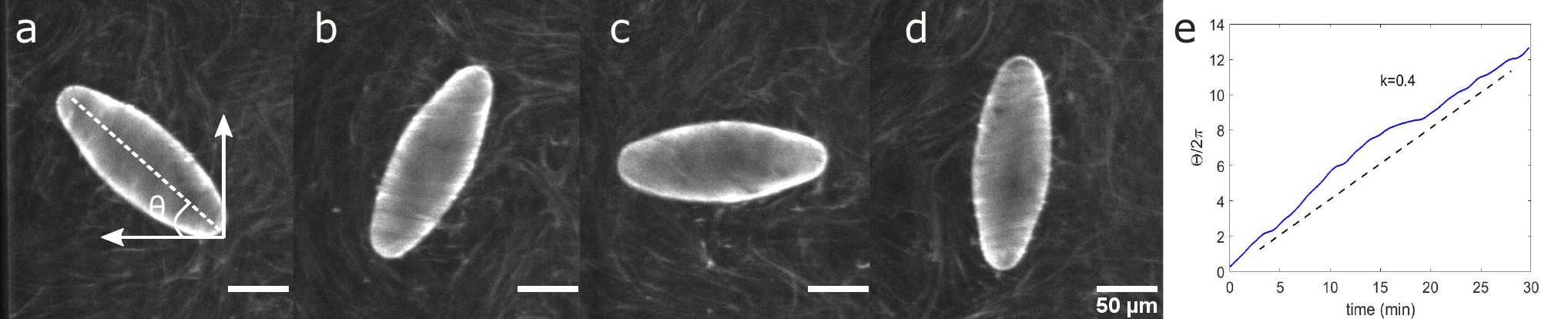}
 \caption{(a)-(d) Fluorescence confocal micrographs of an ellipsoid in the bipolar state showing a solid body rotation; images were taken every 60s. (e) Temporal evolution of the normalized orientation of the major axis of the ellipsoid, showing a constant angular speed.}
 \label{fig_5}
\end{figure*}

\subsubsection*{Effect of external flow fields}

The discrepancies between the experimental and simulation results prompted us to adopt a non-uniform frictional pattern that embodies the complex interplay between the hydrodynamic interactions of the ellipsoidal droplets with the nearby capillary wall and the smectic structure of the inner ellipsoidal droplets. The proposed frictional pattern imposes two favorable paths that are almost mutually perpendicular. Near the poles, preferential lanes orient parallel to the equator, while on the main body, the easy direction is tilted \SI{25}{\degree} with respect to the ellipsoid major axis, see \figref{fig_4}(a). To rationalize this complex pattern, we need to investigate the rigid body behavior of the experimental droplets when they are dispersed in the active bath. \figref{fig_5} shows the temporal evolution of the orientation of the major axis $\theta$ of the droplet with respect to the horizontal axis. Although the confining volume inside the capillary is large compared to the droplet size, our analysis is restricted to two dimensions, since the buoyancy of the droplet keeps the droplet close to the top wall. One can appreciate that the rotation angle $\theta (t)$ increases almost linearly with time for the final dipolar regime. These results suggest that the periodic final dynamics on the active shell is connected to the persistent rotations of the droplet, which arise from hydrodynamic interactions of the droplet with the confining boundary. \par

In order to shed light on these findings, we consider two defect pairs $counter-rotating$ at the poles, while we adjust the boundary conditions close to the surface of the ellipsoid, to reproduce the experimental configuration (\href{run:../SI/Movie_6_rolling.avi}{Movie 8} in SI). The spinning behavior of the defects induces strong shear flows in the thin gap between the droplet and the nearby substrate with opposite signs at the poles. This results in a net hydrodynamic torque that promotes a rigid body rotation of the ellipsoid perpendicularly to its major axis. Interestingly, as the ellipsoid continuously rotates, the surrounding hydrodynamic force couples to the anisotropic friction of the droplet, at the AN/smectic interface, leading to a resultant easy flow pattern that determines the navigation of defects. This hydrodynamic force is stronger at the equatorial region than at the poles, which are farther from the capillary wall, explaining the need to impose a non-uniform effective frictional pattern. \par

The hybrid template shown in \figref{fig_4}(a) successfully replicates the dynamics observed for the final dynamic state in our experiments. Two pairs of $+1/2$ defects periodically oscillate between two regimes, giving rise to $counter-rotating$ vortices at the two opposite ends of the ellipsoid and to a linear translation away from the poles, as shown in \figref{fig_4}(b)-(g). Note that this last configuration originates from the extensile nature of the active stresses, which induce diagonally oriented bending instabilities. Furthermore, the duration of the rotational regime being larger than that of the linear translation, leads to a higher likelihood for spotting defects at the poles, see \figref{fig_4}(i), thereby explaining the propensity of defects for attraction to regions with higher positive Gaussian curvature displayed in \figref{fig_4}(h). The pulse of the defect motion is further regulated by the anisotropic friction of the hybrid pattern which, in close agreement with experimental observations, see \figref{fig_2}(g), amplifies the angular speed during circulatory motion at the poles and minimizes it near the equator of the droplet, as evidenced in \figref{fig_4}(j)-(k). Note that the open circles on \figref{fig_4}(j)-(l) correspond to time series snapshots displayed in \figref{fig_4}(b)-(g). The continual oscillatory behavior between the two regimes can be understood by analyzing the temporal evolution of elastic free energy. At the poles, where defects reach the highest angular velocity, the elastic free energy is minimized. By injecting energy into the system, uniformly aligned active units are subject to bending instabilities, thereby enabling defects to escape from the poles and to travel diagonally toward the mid-plane, where the elastic free energy is maximized, as shown in \figref{fig_4}(l). Notably, the chirality of the defect rotation persists over many periods, despite their continual exchange, in agreement with experiments. The final defect trajectories on AN ellipsoids are shown in \href{run:../ SI/Movie_7_manu32_short.mp4}{Movie 9} in SI and are translated into the spatiotemporal variations of the nematic texture displayed in \href{run:../SI Movie_8_manu32}{Movie 10} in SI. \par


\section{Final remarks and conclusions}

We have presented an experimental system and a theoretical framework that enables to study in detail the dynamics of ANs confined to ellipsoidal smectic surfaces, where Gaussian curvature is non-uniform. The ability of the interfacial material to sense curvature gradients and the friction anisotropy imposed by the smectic substrate, drives periodic oscillations, self-regulated by external hydrodynamic forces that dictate remarkable chiral defect trajectories. Despite the periodic oscillation between a rotational and a translational regime enforced by the ellipsoidal geometry, the frictional anisotropy arising from the smectic texture of the droplet and hydrodynamic interactions with an external wall preserves the chirality of the defects. 

The number of defects on the ellipsoidal surface evolves during the aging of the AN. Initially, the four topologically required $+1/2$ defects are accompanied by additional pairs of $+1/2$ and $-1/2$ defects, which do not alter the total topological charge of the ellipsoid. In the rotational regime, the droplet displays quadrupolar symmetry: the two pairs of topologically required defects occupy the poles of the ellipsoid, while the additional defects form an equatorial belt. As time passes, this initial quadrupolar state is replaced by one with dipolar symmetry, resulting from the suppression of the defect belt. The reduction of the number of defects during the aging of the AN is likely due to an increase in the active length-scale, $l_{\alpha} \propto\sqrt{K/\alpha}$, resulting from the greater stiffness of the AN as the microtubule bundles adsorb at the interface (see Section 1 in SI).

The final dipolar dynamic state is robust against changes in the aspect ratio of the ellipsoid. Although for this study we have focused on ellipsoidal droplets with aspect ratio $c/a \approx 2$, we have also witnessed the emergence of a final dipolar state in ellipsoidal droplets with lower ($c/a \approx 1.5$) and higher ($c/a \approx 3$) aspect ratios. Such an oscillating chiral dipolar state, however, disappears when the smectic droplet has a spherical shape. In that case, the four $+1/2$ defects follow non-regular trajectories and non-periodic dynamics. This behavior seems to stem from the more disordered organization of the smectic layers in the spherical geometry.    

Our findings suggest that ellipsoidal active shells can serve to deepen our understanding of a wide range of biological processes, from tissue morphogenesis to mitosis \cite{maroudas2021, copenhagen2021, novak2018, zhu2017}, which rely on the coupling between defect dynamics and Gaussian curvature, or on the emergence of dipolar symmetry. \par

\section*{Materials and Methods} \label{sec:material}

\subsection*{Production of smectic droplets}

\subsubsection*{Microfluidic device fabrication}

Ellipsoidal droplets were produced from thin double emulsions (W/LC/W) prepared using glass capillary microfluidics \cite{utadaMonodisperseDoubleEmulsions2005a}. Our device is made of two cylindrical capillaries of 1mm outer diameter inserted into the opposite ends of a 1.02 mm square capillary. The two cylindrical capillaries are tapered on one side using a micropipette puller. Their tips are cut with a microforge to reach inner diameters of 60 $\mu$m, for the injection capillary, and 120 $\mu$m, for the collection capillary. The injection capillary, employed to inject the inner phase, is immersed in a solution
containing 0.2v$\%$ n-octadecyl-trimethoxysilane (376213, Sigma), 20v$\%$ chloroform and
79.8v$\%$ hexane, for 5 min to render its surface hydrophobic. Subsequently, the capillary is  washed for 2min with chloroform, dried with compressed air and heated for at least 8 hours at \SI{200}{\celsius}.
The cylindrical capillaries are then inserted facing each other into the square capillary. Their tips are aligned on the microscope, maintaining a separation distance of approximately 60 $\mu$m. Dispensing needles are finally fixed at the three inlets of the device to inject the three phases constituting the double emulsions. \par

\subsubsection*{Production of smectic shells and ellipsoidal droplets} We introduce a 2wt$\%$ Pluronic 127 solution through the injection capillary, and 8CB liquid crystal through the square capillary, which co-flow forming a composed jet at the entrance of the collection cylindrical capillary. The outer solution, composed of Pluronic 127 2wt$\%$ and glycerol 60wt$\%$ in Milli-Q water, is pumped from the other side of the square capillary, flow focusing the composed jet into the collection one. Such composed jet becomes unstable and breaks up into double emulsions. The device is locally heated at a temperature ranging between 33.5$\si{\degree}$C and 40.5$\si{\degree}$C, to maintain the 8CB LC in the nematic phase. We observe a stable dripping regime at typical flow rates of 1000$\mu$L/hr, 400$\mu$L/hr and 7500 $\mu$L/hr for the inner, middle and outer fluids respectively. The shells are then collected in a solution with the same composition as the inner phase, preheated at 60$\si{\degree}$C. In a second step, the collection vial is kept at room temperature for 15 minutes to trigger the N/SmA transition of the LC, inducing the bursting of the shells and the formation of the ellipsoidal droplets, as shown in FIGURE.

\begin{figure}[!htb]
 \centering
 \includegraphics[scale=0.6]{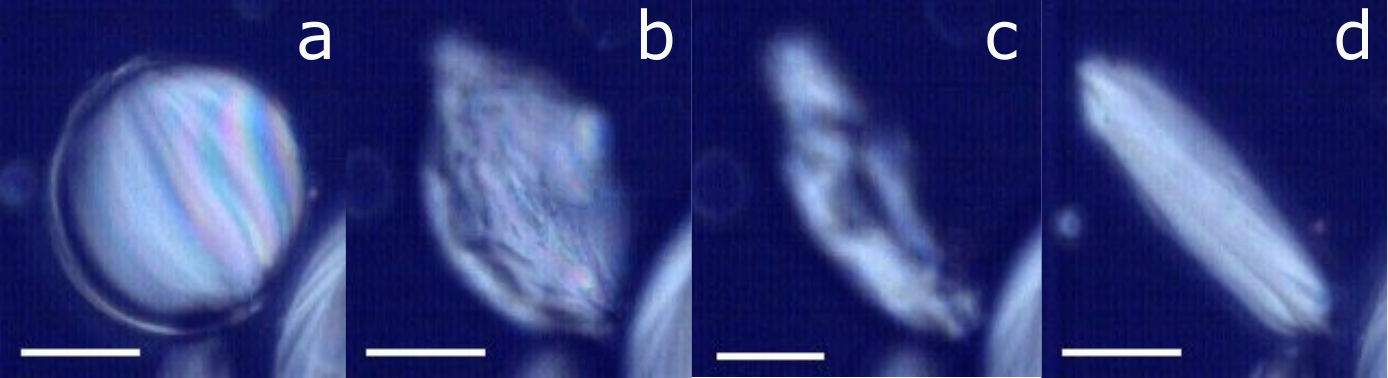}
 \caption{Bursting of a smectic shell and formation of an elongated LC droplet. Scale bar is $50 \mu m$.}
 \label{fig_6}
\end{figure}

\subsection*{Microtubules polymerization}
Polymerization was carried out by incubating at 37$\si{\degree}$C for 30 min, an aqueous
mixture of heterodimeric ($\alpha$,$\beta$)-tubulin from bovine brain (Brandeis University
Biological Materials Facility), M2B buffer (80 mM Pipes, 1 mM EGTA, and 2 mM $MgCl_2$)
(Sigma; P1851, E3889, and M4880, respectively), dithiothrethiol (43815, Sigma) reducing
agent, and a non-hydrolysable guanosine triphosphate (GTP) analogue GMPCPP (guanosine-5\' -
[($\alpha$,$\beta$)-methyleno] triphosphate) (NU-405, Jena Biosciences) that  suppresses the
dynamic instability of MTs. By adjusting GMPCPP concentration, tubulin heterodimers
associate in a controlled way to produce a high-density suspension of short MTs (1–2
$\mu$m). For fluorescence observation, 3$\%$ of the tubulin was labeled with Alexa 647
(A20006, Thermo Fisher Scientific). The final solution was kept at room temperature for 5 h,
frozen in liquid nitrogen and stored at -80$\si{\degree}$C for future use.

\subsection*{Kinesin expression}
In this experiment, heavy-chain kinesin-1 K401-BCCP-6His from Drosophila melanogaster
(truncated at residue 401, fused to biotin carboxyl carrier protein (BCCP), and labeled with
six histidine tags) was expressed in Escherichia coli using the plasmid WC2 from the Gelles
Laboratory (Brandeis University) and purified with a nickel column. After dialysis against
500 mM imidazole buffer, the concentration of the suspension was adjusted to $60 \%$ (wt/v)
with a sucrose solution and estimated by means of absorption spectroscopy. The final protein
solution was stored at \SI{-80}{\celsius} until used.

\subsection*{Active solution preparation}
Kinesin/streptavidin motors clusters were prepared by incubating biotinylated kinesins with
tetrameric streptavidin (43-4301, Invitrogen) at a 2:1 stochiometric ratio, for 30 min,
under the reducing action of 2$\mu$M DTT. The resulting suspension was mixed with a standard
solution containing the M2B buffer, and a PEG (20 kDa; 95172, Sigma) depleting agent. To
prevent protein denaturation and photobleaching during the fluorescence acquisition, 5,8 mM
DTT, 0.2 mg/ml glucose oxidase, 0.04 mg/ml catalase, 2.1 mM Trolox and 3.5 mg/ml glucose
antioxidants were incorporated to the mixture. The activity of the system was provided by
chemical energy in the form of ATP (A2383, Sigma) (1.5 mM) and was constantly regenerated by
the action of phosphoenolpyruvate (PEP) (P7127, Sigma) and pyruvate kinase/lactate
dehydrogenase (PK/LDH) (434301, Invitrogen).

\subsection*{Sample preparation}
The active gel was prepared by introducing 1$\mu$L of microtubule suspension in 5$\mu$L of active
solution. Biocompatility and microtubule adsorption at the water/oil interface, in subsequent steps,
were facilitated by the addition of 1$\mu$L Pluronic F-127 (P2443, Sigma) 17wt$\%$ and
1$\mu$L Tween-80 (P1754, Sigma) 17wt$\%$ surfactants to the previous solution. Afterward,
1$\mu$L of the smectic droplets suspension was carefully incorporated. The final solution
was gently introduced into a 0.6 mm square capillary, previously treated with an acrylamide
brush. The latter coating was prepared by following reported protocols\cite{lau2009} to
prevent the adhesion of the active material to the glass walls. The use of a square capillary of such dimensions has several advantages over a standard flow cell \cite{guillamat2018}. First, it allows us to manipulate small amounts of the active gel and to concentrate the droplets in a small volume. Second, the chemical energy (ATP) of the active bath can be renewed as often as needed since one of the capillary ends is kept open. Lastly, in the capillary, the evaporation of the active solution is reduced with respect to a standard flow cell. 

\subsection*{Imaging}
Active nematics observation, based on the fluorescence of labeled microtubules, was performed using a spinning disk confocal on a Nikon TI-E inverted microscope equipped with a Perfect Focus System (PFS) for continuous maintenance of focus. Images were typically captured every 500ms with a 10x objective and an Andor Zyla 4.2MP camera operated with NiS-Elements software. With the x10 objective, the confocal mode is not perfectly reached, giving rise to a depth of field of tens of microns, which allows us to focus the whole droplet hemisphere, a few microns in depth. Images for the 3D reconstruction of defect trajectories (\figref{fig_1}(g), \figref{fig_2}(f) and Movie 3 in SI) were acquired by simultaneously capturing 4 different planes, two of the top and two of the bottom planes of the droplet, at a rate of 0.7 frames/s. 

\subsection*{Image analysis}

Defect tracking was performed from confocal z-projection images, using the $Manual$ $Tracking$ Fiji/imageJ plugin. For each frame of a movie, defects were identified as dark spots with comet-like shape. 2D Trajectories were further computed  with a home-made MATLAB program. The angular speed of defects was calculated from their 3D trajectories, reconstructed by projecting 2D trajectories onto the ellipsoidal surface.

\section*{Simulation details} \label{sec:simulation}
The systems considered here are described in terms of a nematic tensorial order
parameter $\bf{Q}$ and a flow field $\bf{u}$. The standard theory of active nematodynamics
is used to quantify their spatiotemporal evolution. For uniaxial systems, the nematic order
parameter is written in the form ${\bf{Q}} = S({\bf{nn}} - {\bf{I}}/3)$ where unit vector
$\bf{n}$ is the nematic director field and $S$ is an order parameter that measures the
extent of orientational ordering. The evolution of this non-conserved order parameter obeys the
strongly non-linear equation 
\begin{equation}
 (\frac{\partial }{{\partial t}} + {\bf{u}} \cdot {\bm{\nabla}} ){\bf{Q}} - {\bf{S}} = \Gamma
 {\bf{H}},
 \label{beris}
\end{equation}
where the advection term is generalized by
${\bf{S}} = (\xi {\bf{A}} + {\bf{\Omega }}) \cdot ({\bf{Q}} + \frac{{\bf{I}}}{3}) +
({\bf{Q}} + \frac{{\bf{I}}}{3}) \cdot (\xi {\bf{A}} - {\bf{\Omega }})
-2\xi ({\bf{Q}} + \frac{{\bf{I}}}{3})({\bf{Q}}:{\bm{\nabla}} {\bf{u}})$,
which accounts for the response of the nematic order parameter to the symmetric, ${\bf{A}}$,
and antisymmetric, ${\bf{\Omega}}$, parts of the velocity gradient tensor (${\bm{\nabla}}
{\bf{u}}$). Here, $\xi$, is the flow aligning parameter, chosen to be $\xi = 0.7$ for flow
aligning elongated units.
The molecular field ${\bf{H}} =  - (\frac{{\delta {\mathcal{F}_{\textrm{LdG}}}}}{{\delta
 {\bf{Q}}}} -
\frac{{\bf{I}}}{3}\Tr\frac{{\delta {\mathcal{F}_{\textrm{LdG}}}}}{{\delta {\bf{Q}}}})$
embodies the relaxational dynamics of the nematic, which drives the system toward the
 configuration with minimum Landau-de Gennes free energy
\begin{equation}
 \mathcal{F}_{\textrm{LdG}} = \int_V {f_{\textrm{LdG}} \; \textrm{d}V}.
\end{equation} 

The free energy density, $f_{\textrm{LdG}}$, is the sum of bulk and elastic energies given
by:
\begin{eqnarray}
 {f_{\textrm{LdG}}}\; = && \; \frac{{{A_0}}}{2}(1 -
 \frac{U}{3})\;\Tr({{\bf{Q}}^2})\; - \;\frac{{{A_0}U}}{3}\;\Tr({{\bf{Q}}^3})\; \nonumber\\
 && + \frac{{{A_0}U}}{4}\;{(\Tr({{\bf{Q}}^2}))^2} + \frac{L}{2}{({\bm{\nabla}} {\bf{Q}})^2}.
\end{eqnarray}
The relaxation rate is controlled by the collective rotational diffusion constant $\Gamma$.
The phenomenological coefficient $A_0$ sets the energy scale, $U$ controls the magnitude of
the order parameter, and $L$ is the elastic constant in the one-elastic constant
approximation.
At the boundary surface with unit normal $\bm{\nu}$, the anchoring condition is imposed by
adding a surface term to the free energy,
$\mathcal{F}_{\textrm{surf}} = \int_{\partial V} {f_{\textrm{surf}}\; \textrm{d}S}$.
The fourth order Fournier-Galatola free energy density is adopted to apply the non-degenerate
planar anchoring boundary condition
\begin{equation}
 {f_{{\rm{surf}}}} = \frac{1}{2}{\cal W}{({\bf{\bar Q}} - {{{\bf{\bar Q}}}_ \bot })^2} +
 \frac{1}{4}{\cal W}{({\bf{\bar Q}}:{\bf{\bar Q}} - {S^2})^2},
 \label{fournier}
\end{equation}
where $\mathcal{W}$ controls the anchoring strength,
${\bf{\bar Q}} = {\bf{Q}} + \frac{1}{3}S_{eq}{\bm{\delta}}$,
its projection to the surface
${{{\bf{\bar Q}}}_ \bot } = {\bf{p}} \cdot {\bf{\bar Q}} \cdot {\bf{p}}$, and
${\bf{p}} = {\bm{\delta}}  - {\bm{\nu \nu}}$.  \par

The local fluid density $\rho$ and velocity $\bf{u}$ are governed by the generalized
incompressible Navier-Stokes equations, modified by a frictional dissipative term
\begin{align}
 \rho (\frac{\partial }{{\partial t}} + {\bf{u}} \cdot {\bm{\nabla}} ){\bf{u}} & =
 {\bm{\nabla}}  \cdot {\bm{\Pi}} - \bf{f}\bf{u}.
 \label{navier2}
\end{align}
The total asymmetric stress tensor ${\bf \Pi}={\bf \Pi} ^p+{\bf \Pi} ^a$ is a sum of a
passive and an active stress, and $\bf{f}$ is the diagonal tensor describing frictional
damping between the nematic fluid and the underlying substrate. The viscoelastic
properties of the nematic are lumped in the passive stress, which is sum of the viscous and
elastic terms
\begin{eqnarray}
 {{\bm{\Pi}} ^p}  && = 2\eta {\bf{A}} - P_0{\bf{I}} + 2\xi ({\bf{Q}} +
 \frac{{\bf{I}}}{3})({\bf{Q}}:{\bf{H}}) - \xi {\bf{H}} \cdot ({\bf{Q}} + \frac{{\bf{I}}}{3})
 \nonumber \\
 && - \xi ({\bf{Q}} + \frac{{\bf{I}}}{3}) \cdot {\bf{H}} - {\bm{\nabla}}
 {\bf{Q}}:\frac{{\delta
  {\mathcal{F}_{\textrm{LdG}}}}}{{\delta {\bm{\nabla}} {\bf{Q}}}} + {\bf{Q}} \cdot {\bf{H}} -
 {\bf{H}} \cdot {\bf{Q}},
\end{eqnarray}
and the active stress
\begin{equation}
 {{\bm{\Pi}} ^a} =  - \zeta {\bf{Q}}.
\end{equation} 

Here, $\eta$ is the isotropic viscosity, $P_0$ is the isotropic bulk pressure, and
$\zeta$ measures the activity strength. A flow is generated when $\bf{Q}$ has spatial
gradient with $\zeta > 0$ for extensile systems and $\zeta < 0$ for contractile ones.
We employ a hybrid lattice Boltzmann method to solve the coupled governing partial
differential equations (Eqs.~\ref{beris}, \ref{navier2}) \cite{marenduzzo2007, ravnik2013, zhang2018, sokolov2019}. 
The time integration is performed using an Euler
forward scheme; the spatial derivatives are carried out using a second order central
difference and the coupling is enforced by exchanging local fields between these algorithms
at each time step. Simulations were performed on a three-dimensional lattice where the
active layer were confined between two ellipsoid with a uniform shell of thickness $4$ lattice
spacing and the inner prolate spheroid's semi-minor and semi-major axes were chosen to be
$20,\, 20,\, 40$ respectively, giving the aspect ratio of the inner shell to be slightly smaller than $2$ (uniform thickness of
the shell results in a different aspect ratio for the outer shell, slightly smaller than $2$). All dimensions are
expressed in lattice  units. The medium viscosity was set to $\eta = 1/6$, and the
collective rotational viscosity to $\Gamma = 0.3$. We chose the following parameters
throughout the simulation $A_0 = 0.1$, $L = 0.1$, $U = 3.0$ (giving $S=0.5$). Planar
anchoring conditions with strength $\mathcal{W} = 0.1$ and a no-slip velocity field at the
surface of the inner and outer ellipsoids were enforced. The system was initialized with the
director field tangentially oriented along the major axis. The thin shell confined between
the ellipsoids was activated by applying uniform extensile active stresses to the nematic.

\begin{acknowledgments}
This project has received funding from the European Union's Horizon 2020 research and innovation program, under the Marie Sklodowska-Curie grant agreement No 754387, and from the French National Research Agency, under grant ANR-18-CE09-0028-02. The theoretical work was funded by the National Science Foundation, through the University of Chicago MRSEC. J.I.-M., and F.S. acknowledge funding from MICINN/AEI/10.13039/501100011033 (Grant No. PID2019-108842GB-C22). We thank B. Martínez-Prat, M. Pons, A. LeRoux, and G. Iruela (Universitat de Barcelona) for their assistance in the expression of motor proteins. We thank Brandeis University MRSEC Biosynthesis facility (supported by NSF MRSEC 2011846) for providing the tubulin.
\end{acknowledgments}

\setcounter{figure}{0}
\renewcommand{\thefigure}{S\arabic{figure}}%


\onecolumngrid
\begin{center}
\section*{\centering \large \textbf{Dynamics of active defects on the anisotropic surface of an ellipsoidal droplet \\
Supplementary Information}} 
\vspace{0.5cm}
{\normalsize Martina Clairand$^{1,*}$, Ali Mozaffari$^{2,3,*}$, Jer\^ome Hardo\"uin$^{4,5}$, Rui Zhang$^{2,6}$, Claire Dor\'e$^{1}$, Jordi Ign\'es-Mullol$^{4,5}$, Francesc Sagu\'es$^{4,5}$, Juan J. de Pablo$^{2,7}$, Teresa Lopez-Leon$^{1}$} \\
\vspace{0.2cm}
{\normalsize \emph{$^1$ Laboratoire Gulliver, UMR CNRS 7083, ESPCI Paris, PSL Research University, 75005 Paris, France}} \\
{\normalsize \emph{$^2$ Pritzker School of Molecular Engineering, The University of Chicago, Chicago, Illinois
 60637, USA}} \\
{\normalsize \emph{$^3$ OpenEye Scientific, Cadence Molecular Sciences, Boston, Massachusetts 02114, USA}} \\
{\normalsize \emph{$^4$ Departament de Qu\'{\i}mica F\'{\i}sica, Universitat de Barcelona, 08028 Barcelona, Spain}} \\
{\normalsize \emph{$^5$ Institute of Nanoscience and Nanotechnology, Universitat de Barcelona, 08028 Barcelona, Spain}} \\
{\normalsize \emph{$^6$Department of Physics, Hong Kong University of Science and Technology, Clear Water Bay, Kowloon, Hong Kong}} \\
{\normalsize \emph{$^7$Center for Molecular Engineering, Argonne National Laboratory, Lemont, Illinois
 60439, USA}}\\
{\normalsize{ (Dated: \today})}\\
\vspace{0.5cm}

\end{center}
\twocolumngrid
%


\section*{Oscillatory dynamics of topological defects in the final dipolar state} 

\begin{figure*}[!htp]
\centering
\includegraphics[width=1\textwidth]{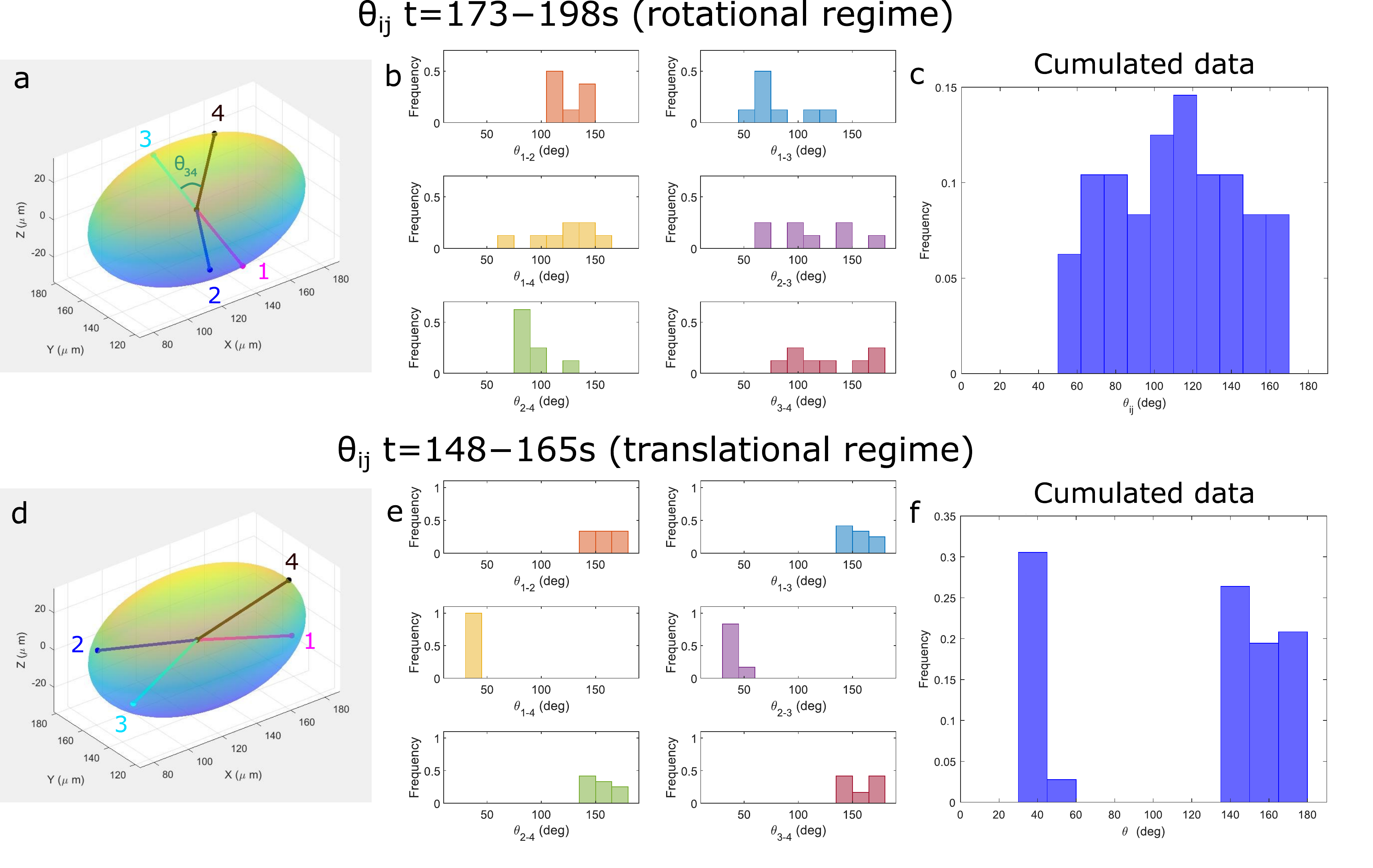}
\caption{\textbf{Distributions of the relative central angles between pairs of defects $\theta_{ij}$} during (a-c) the translational regime, (d-f) the rotational regime.}
 \label{fig_s3}
\end{figure*}

To visualize the oscillatory dynamics of the AN flows in the final dipolar dynamical state, we have reconstructed the 3D-trajectory of the topological defects at the surface of an ellipsoid (see Fig.2 (f) of the main text, Movie 3 and the left panel in Movie 4 in SI). The representations were obtained by tracking simultaneously the defects on the top and bottom planes of the droplet (left panel of Movie 3 in SI), and projecting the resulting 2D trajectories on each of the two hemi-ellipsoidal surfaces, via a custom MATLAB program. As shown in Movie 3 and Movie 4, oscillations appear as periodic cycles of defect rotation at the poles and translation from pole to pole. The defect configuration of each regime is well-characterized by the distribution of $\theta_{ij}$, which refers to the central angular distance between defect pairs (\figref{fig_s3}). During the translational regime (\figref{fig_s3} (a-c)), the distribution is homogeneous, with $\theta_{ij}$ comprised between $50\degree$ and $170\degree$ and a maximal occurrence near $110\degree$. In the rotational regime (\figref{fig_s3} (d-f)), on the other hand, defect pairs near the poles are separated by the angles $\theta_{14}$ and $\theta_{23}$ , with values below $60\degree$, while opposite pairs of defects form angles $\theta_{12}$ ,$\theta_{13}$, $\theta_{24}$ and $\theta_{34}$, above $140\degree$.

\section*{Simulation of active nematics on a 2D planar interface with frictional forces}

\begin{figure*}[!htb]
 \centering
 \includegraphics[width=\textwidth]{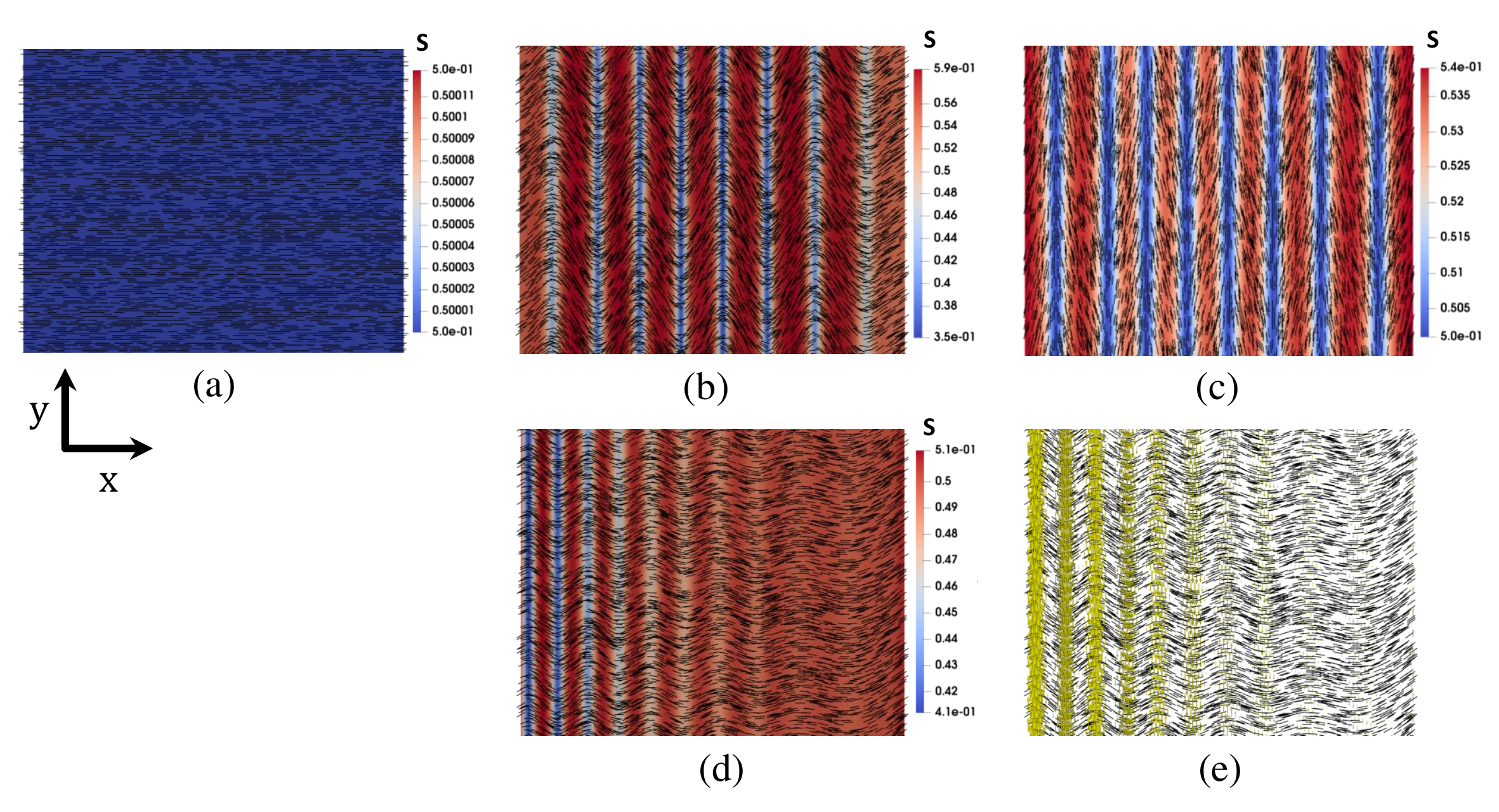}
 \captionsetup{width=\textwidth}
 \caption{\textbf{The effect of friction on the director field and velocity profile of an active nematic layer confined on a 2D \ planar interface.}
 Simulations were performed on a $200 \times 200$ two-dimensional lattice. The medium viscosity was set to $\eta = 1/6$, and the collective rotational viscosity to $\Gamma = 0.3$. The following parameters also were chosen $A_0 = 0.05$, $L = 0.05$, $U = 3.0$ (giving $S=0.5$). (a) The system was initialized with the director field oriented along the $x$-axis and this was the starting point for all simulations. (b-c) Non-uniform frictional forces are applied with the only non zero coefficient of damping matrix being $\delta_{xx}$ ($f_{0} = 0.3$ and $\zeta = 0.005)$. (d-e) A gradient of friction is applied, which grows linearly along the $x$-axis (${\bf{f}} = {f_0} {\bf{I}}$, $f_{0} = 0.2(1+x/L_{x})$ and $\zeta = 0.012)$. A Poiseuille-like flow along the stripes are presented in (e) with $|v_{max}| \approx 2.95\times10^{-3}$ in simulation units at the low friction region. Yellow arrows represent the local velocity field, with the magnitude of the velocity being normalized with its maximum value along the lines of high elastic distortions.}
 \label{fig_s4}
\end{figure*}

We use numerical simulations to investigate how \textit{frictional forces} control the dynamics of active nematics confined to a 2D planar interface. Initially, the director field is aligned along the $x$-axis, as shown in \figref{fig_s4} (a). Once the extensile active force is applied on the system, bending instabilities yield the formation of topological defects and chaotic flows. However, the presence of frictional forces leads to regular spatio-temporal patterns in the flow and director fields, as shown in the remaining panels of \figref{fig_s4}. 

We first examined the effect of incorporating an anisotropic frictional pattern where only the $\delta_{xx}$ component of the tensorial damping term in Eq.5 is non-zero. This setup penalizes the flows parallel to the initial orientation of the director field ($x$-axis) and provides an easy flow direction along the $y$-axis. Initially, the system shows bending instabilities similar to those in the previous case, see \figref{fig_s4} (b). However, due to the large elastic distortions and high shear stresses generated as a result of the formation of these elastic bands, at later times, the director undergoes additional distortions. The system finds it more energetically favorable to unbind pairs of $\pm 1/2$ defects and to form a stable nematic pattern with dominant splay deformation. The frictional forces imposed along the $x$-axis impedes the growth of additional bending instabilities and the nematic structure remains stable, as shown in \figref{fig_s4} (c). This result is closely related to the experimental observation made by Guillamat \textit{et al} in a microtubule-kinesin active nematic in contact with 8CB liquid crystal aligned via a magnetic field, where the high friction anisotropy imposed by the smectic layer led to the formation of anti-parallel flowing lanes \cite{guillamat2016}. 

We then consider a case where the substrate in contact with the active layer does not impose a directional frictional anisotropy but a gradient in the friction magnitude. Here, the strength of the frictional damping force, $f_0$ grows linearly along the $x$-axis. This scenario can be achieved by having the active layer in contact with a substrate of varying viscosity.  \figref{fig_s4} (d-e) shows that lanes of anti-parallel flow develop with bands of different width. At the higher value of friction strength (higher values of $x$ in \figref{fig_s4} (d-e)), the suppression of momentum propagation results in bands with a larger width along with smaller values of velocity. \par 
With these examples we have demonstrated that by controlling the friction at a substrate, the otherwise chaotic active nematic flows can be tamed to precisely sculpt well-ordered spatio-temporal states and flow fields.
\par



\clearpage


%


\end{document}